\documentclass[amsmath,aps,prl,twocolumn,showpacs,superscriptaddress,groupedaddress]{revtex4}  
\usepackage{caption}
\usepackage{CJK}
\usepackage{graphicx}  
\usepackage{dcolumn}   
\usepackage{amssymb}   
\usepackage[utf8]{inputenc}
\usepackage{lscape}
\usepackage{booktabs}
\usepackage{mathptmx, courier, pifont}
\usepackage[scaled=0.92]{helvet}
\usepackage[T1]{fontenc}
\usepackage{textcomp}
\usepackage{rotating}
\usepackage{verbatim}
\usepackage{picinpar}
\usepackage{wrapfig}
\usepackage{multirow}
\usepackage[normalem]{ulem}
\usepackage[colorlinks=true, citecolor=blue, urlcolor=blue, linkcolor=blue,breaklinks=true]{hyperref}
\usepackage{orcidlink}
\usepackage{flushend}
\PassOptionsToPackage{unicode}{hyperref}
\PassOptionsToPackage{naturalnames}{hyperref}
\usepackage{parskip} 

\begin{document}

\title{Simultaneous impacts of nuclear shell structure and collectivity on $\beta$ decay: Evidence from $^{80}$Ga$_{49}$}

\author{{R. Li}\orcidlink{0000-0002-2782-2333}}
\email[Contact author: ]{liren824@gmail.com}  
\affiliation{Université Paris-Saclay, CNRS/IN2P3, IJCLab, 91405 Orsay, France}
\affiliation{KU Leuven, Instituut voor Kern- en Stralingsfysica, Celestijnenlaan 200D, B-3001 Leuven, Belgium}

\author{{D. Verney}\orcidlink{0000-0001-7924-2851}}
\affiliation{Université Paris-Saclay, CNRS/IN2P3, IJCLab, 91405 Orsay, France}
\author{{G.~De Gregorio}\orcidlink{0000-0003-0253-915X}}
\affiliation{Dipartimento di Matematica e Fisica, Universit\'a degli Studi della Campania "Luigi Vanvitelli", Viale Abramo Lincoln 5, I-81100 Caserta, Italy}
\affiliation{Istituto Nazionale di Fisica Nucleare, Complesso Universitario di Monte S. Angelo, Via Cintia, I-80126 Napoli, Italy}
\author{R.~Mancino}
\altaffiliation[Present address: ]{Institute of Particle and Nuclear Physics, Faculty of Mathematics and Physics, Charles University, V Holešovičkách 2, 180 00 Prague, Czech Republic.}
\affiliation{Institut f{\"u}r Kernphysik (Theoriezentrum), Fachbereich Physik, Technische Universit{\"a}t Darmstadt, Schlossgartenstrasse 2, 64298 Darmstadt, Germany}
\affiliation{GSI Helmholtzzentrum f{\"u}r Schwerionenforschung, Planckstrasse 1, 64291 Darmstadt, Germany}
\author{I. Matea}
\affiliation{Université Paris-Saclay, CNRS/IN2P3, IJCLab, 91405 Orsay, France}
\author{{L.~Coraggio}\orcidlink{0000-0002-4327-9107}}
\affiliation{Dipartimento di Matematica e Fisica, Universit\'a degli Studi della Campania "Luigi Vanvitelli",Viale Abramo Lincoln 5, I-81100 Caserta, Italy}
\affiliation{Istituto Nazionale di Fisica Nucleare, Complesso Universitario di Monte S. Angelo, Via Cintia, I-80126 Napoli, Italy}
\author{{N.~Itaco}\orcidlink{0000-0002-9508-2613}}
\affiliation{Dipartimento di Matematica e Fisica, Universit\'a degli Studi della Campania "Luigi Vanvitelli",Viale Abramo Lincoln 5, I-81100 Caserta, Italy}
\affiliation{Istituto Nazionale di Fisica Nucleare, Complesso Universitario di Monte S. Angelo, Via Cintia, I-80126 Napoli, Italy}
\author{{M. N. Harakeh}\orcidlink{0000-0002-7271-1712}}
\affiliation{ESRIG, University of Groningen, Zernikelaan 25, 9747 AA Groningen, The Netherlands}
\author{{C. Delafosse}\orcidlink{0000-0001-5717-2426}}
\affiliation{Université Paris-Saclay, CNRS/IN2P3, IJCLab, 91405 Orsay, France}
\affiliation{Department of Physics, Accelerator Laboratory, P.O. Box 35, University of Jyväskylä, FI-40014 Finland}
\author{{F. Didierjean}\orcidlink{0009-0002-4950-3162}}
\affiliation{Institut Pluridisciplinaire Hubert Curien, CNRS/IN2P3 and Université de Strasbourg, Strasbourg, France}

\author{{L. A. Ayoubi}\orcidlink{0000-0002-1441-9094}}
\affiliation{Université Paris-Saclay, CNRS/IN2P3, IJCLab, 91405 Orsay, France}
\affiliation{Department of Physics, Accelerator Laboratory, P.O. Box 35, University of Jyväskylä, FI-40014 Finland}
\author{H. Al Falou}
\affiliation{Faculty of Sciences 3, Lebanese University, Michel Slayman Tripoli Campus, Ras Maska 1352, Lebanon}
\author{G. Benzoni}
\affiliation{INFN, Sezione di Milano, Dipartamiento di Fisica, Milano, Italy}
\author{{F. Le Blanc}\orcidlink{0000-0001-6463-289X}}
\affiliation{Université Paris-Saclay, CNRS/IN2P3, IJCLab, 91405 Orsay, France}
\author{{V. Bozkurt}\orcidlink{0000-0003-4651-0447}}
\affiliation{Nigde University, Science Faculty, Department of Physics, Nigde, Turkey}
\author{M. Ciemała}
\affiliation{Institute of Nuclear Physics, Polish Academy of Sciences, Krakow, Poland}
\author{{I. Deloncle}\orcidlink{0000-0002-7646-5117}}
\affiliation{Université Paris-Saclay, CNRS/IN2P3, IJCLab, 91405 Orsay, France}
\author{M. Fallot}
\affiliation{Subatech, CNRS/IN2P3, Nantes, EMN, F-44307, Nantes, France}
\author{C. Gaulard}
\affiliation{Université Paris-Saclay, CNRS/IN2P3, IJCLab, 91405 Orsay, France}
\author{A. Gottardo}
\affiliation{Laboratori Nazionali di Legnaro, I-35020 Legnaro, Italy}
\author{V. Guadilla}
\altaffiliation[Present address: ]{Faculty of Physics, University of Warsaw, 02-093 Warsaw, Poland.}
\affiliation{Subatech, CNRS/IN2P3, Nantes, EMN, F-44307 Nantes, France}
\author{{J. Guillot}\orcidlink{0000-0001-9650-0613}}
\affiliation{Université Paris-Saclay, CNRS/IN2P3, IJCLab, 91405 Orsay, France}
\author{K. Hadyńska-Klęk}
\affiliation{Department of Physics, University of Oslo, Oslo, Norway}
\author{F. Ibrahim}
\affiliation{Université Paris-Saclay, CNRS/IN2P3, IJCLab, 91405 Orsay, France}
\author{N. Jovancevic}
\altaffiliation[Present address: ]{University of Novi Sad, Faculty of Science, Novi Sad, Serbia.}
\affiliation{Université Paris-Saclay, CNRS/IN2P3, IJCLab, 91405 Orsay, France}
\author{{A. Kankainen}\orcidlink{0000-0003-1082-7602}}
\affiliation{Department of Physics, Accelerator Laboratory, P.O. Box 35, University of Jyväskylä, FI-40014 Finland}
\author{M. Lebois }
\affiliation{Université Paris-Saclay, CNRS/IN2P3, IJCLab, 91405 Orsay, France}
\author{{T. Mart\'{i}nez}\orcidlink{0000-0002-0683-5506}}
\affiliation{Centro de Investigaciones Energeticas Medioambientales y Tecnológicas (CIEMAT), Madrid, Spain}
\author{{P. Napiorkowski}\orcidlink{0000-0001-5940-2555}}
\affiliation{Heavy Ion Laboratory, University of Warsaw, 02-093 Warsaw, Poland}
\author{B. Roussiere}
\affiliation{Université Paris-Saclay, CNRS/IN2P3, IJCLab, 91405 Orsay, France}
\author{{Yu. G. Sobolev}\orcidlink{0000-0002-5615-0468}}
\affiliation{Joint Institute for Nuclear Research, Dubna, Russia}
\author{{I. Stefan}\orcidlink{0000-0001-7923-8908}}
\affiliation{Université Paris-Saclay, CNRS/IN2P3, IJCLab, 91405 Orsay, France}
\author{{S. Stukalov}\orcidlink{0000-0003-2948-1947}}
\affiliation{Joint Institute for Nuclear Research, Dubna, Russia}
\author{D. Thisse}
\affiliation{Université Paris-Saclay, CNRS/IN2P3, IJCLab, 91405 Orsay, France}
\author{G. Tocabens}
\affiliation{Université Paris-Saclay, CNRS/IN2P3, IJCLab, 91405 Orsay, France}

\date{\today}

\begin{abstract}

The Gamow-Teller strength distribution covering the entire $\beta$-decay window, up to 10.312(4) MeV, of $^{80g+m}$Ga was measured for the first time in photo fission of UC$_x$ induced by a 50 MeV electron beam. The new data show significant enhancement in the high-energy region with a jump structure. Simultaneously, the $\gamma$ deexciting behavior of $\beta$-populated states presents a competition between deexcitation to 2$_1^+$ [$\beta_2$ = 0.155(9)] and to 2$_2^+$ [$\beta_2$ = 0.053$_{0.009}^{0.008}$)] in $^{80}$Ge. To understand these data, we performed a realistic shell-model calculation and systematic analysis of log $\it{ft}$ ratios between precursors' $\beta$ decay to 2$_2^+$ and to 2$_1^+$ of Ga isotopes. We conclude that these phenomena evidence simultaneous impacts of nuclear shell structure and collectivity on $\it{B}$(GT) distribution and therefore the half-life of the precursor.

\end{abstract}
\maketitle

\section{I. Introduction}
$\beta$ decay keeps its mystery to some extent nowadays, more than 120 years since its discovery in the nuclear medium, which is a self-organized many-body quantum system dominated by the strong interaction. This arises from complexity of the weak-interaction process in nuclei and the response of the nuclei following $\beta$ transitions. From classical Fermi current-current interaction theory, phenomenal effective field theory based on point like interaction hypothesis, nuclear $\beta$-decay is a process of conversion of a neutron to a proton or vice versa. Consequently, it generates a one-particle-one-hole (1p1h) excitation in daughter nuclei. Based on this theory, observables in nuclear $\beta$ decay, mainly half-lives, can be reproduced approximately.

However, recent studies further reveal that the $\beta$ decay of atomic nuclei is not as simple as introduced above. This discovery was triggered, from the theoretical side, by precisely reproducing the $\it{B}$(GT) without using a phenomenological quenching factor \citep{Engel_2017,Suhonen2017,SUHONEN2013153,BROWN1985347,Pinedo1996} either through adding 2p2h correlations in the final-state wave function \cite{gambacurta2020gamow}, or adding nuclear collective vibration in the final-state wave function \cite{niu2015particle}, or including the two-body current contributions of the axial current in the $\beta$-decay operator \cite{gysbers2019discrepancy,PhysRevC.109.014301}. In all cases, multi correlated excitation states must be populated instead of single-particle states. From the experimental side, studies of $\it{B}$(GT) distributions using the total absorption gamma spectroscopy (TAS) technique \cite{ALGORA2021136438,Briz2015,Algora2004} indicate the influence of nuclear deformation on $\beta$ decay, e.g., $^{148}$Tb, $^{150}$Ho, $^{152}$Tm \cite{PhysRevC.93.014308}, and $^{148}$Dy \cite{PhysRevC.70.064301} in the region of $^{146}$Gd; $^{97}$Ag \cite{PhysRevC.60.024315}, $^{100}$Ag \cite{batist1995gamow}, and $^{102}$In \cite{GIERLIK2003313} near $^{100}$Sn.

In this article, we report on the evidence of simultaneous impacts of nuclear shell structure and collectivity on $\beta$-decay properties via (1) investigating the evolution of the cumulated $\it{B}$(GT) jump-structure with the excitation energy from both ground and isomeric states of $^{80}$Ga; (2) comparison with theoretical results obtained within the framework of the realistic nuclear shell model without using a phenomenological quenching factor of the axial coupling constant; (3) analyzing the decay patterns of excited states in the daughter nucleus; and (4) trying to search for correlation between quadrupole deformation of the precursor and selectivity of $\beta$ population. In fact, one expects that if a given precursor has a large quadrupole deformation then it will have a higher probability of decaying to a collective state with higher deformation. This is the case with $^{80m}$Ga with spin-parity of 3$^-$. The electric-quadrupole moment of this isomer is 36.7(16) $\it{e}$ fm$^2$ determined using the laser spectroscopy technique \cite{PhysRevC.82.051302,PhysRevC.96.044324} via fitting hyperfine structure peaks with the $\chi^2$-minimization method. This isomer is more likely to decay to the collective 2$_1^+$ state in $^{80}$Ge than to the 2$_2^+$ state with almost a spherical shape. Consequently, there is a lower log $\it{ft}$ value for decay to 2$_1^+$.

$^{80}$Ga decays to $^{80}$Ge, a nucleus that has $\it{Z}$ = 32 protons and two neutron holes in the closed shell $\it{N}$ = 50. This decay has several unique characteristics that allow one to investigate their impacts. First, the neutron-rich nucleus $^{80}$Ga has a large $\it{Q}_{\beta}$ value with 10.312(4) MeV \cite{wang2017ame2016} so that it offers a wide energy window to observe the excitation spectrum of $^{80}$Ge. Second, the structure of $^{80}$Ge is considered to be dominated by strong shell effects \cite{iwasaki2008persistence}. However, quadrupole including triaxial deformations have been assigned to ground and excited states of $^{80}$Ge via comparison with shell-model calculations using JUN45 and JJ4B interactions \cite{verney2013structure} and via $\it{B(E2)}$ measurements \cite{padilla2005b,iwasaki2008persistence,rhodes2022evolution} which show $\beta_2$ = 0.183(5) that is the error-weighted average from three experiments. Therefore, it is a good case to investigate the role of nuclear collectivity in $\beta$ decay particularly in the closed-shell region.

\section{II. Experiment}

The experiment was performed at the Accelerator Linear and Tandem at Orsay (ALTO) \cite{ibrahim2007alto}. A radioactive $^{80g+m}$Ga ion beam was produced at the ALTO-ISOL facility using photo fission of UC$_x$ induced by a 50 MeV electron beam with an intensity of $\approx$7 $\mu$A. The purification of the beam was obtained in two steps: laser ionization and mass separation. Since at around $\it{A}$ = 80 the only surface-ionized component of a photo-fission generated ion beam is Ga, complete isotope purity was achieved, without any contamination of $^{80}$Rb. This cleanness was tested by the $\beta$-gated $\gamma$ spectrum. No contaminating $\gamma$ rays, such as the $\beta^+$/$\epsilon$-delayed 616.7 keV $\gamma$ line of $^{80}$Kr, were found in the spectrum other than $\gamma$ rays from $^{80}$Ge and its daughter and granddaughter nuclei. The purified $^{80}$Ga beam was directed and implanted on a periodically moving tape to minimize the daughter's activity. The time settings were 0.5 s for background, 5 s for ion collection, and 5 s for decay measurement. The implantation rate was $\approx$10$^4$ pps.

The emitted radiation was detected in the BEta Decay studies at Orsay (BEDO) set up. It was mounted with one cylindrical plastic detector for $\beta$ tagging, surrounded by two high-purity germanium (HPGe) detectors and three PARIS (Photon Array for studies with Radioactive Ion and Stable beams) clusters \cite{ciemala2009measurements,ghosh2016characterization}. The high energy resolution of the HPGe in the 0$-$6 MeV energy range makes the $\gamma$-$\gamma$ coincidence technique very effective not only to reconstruct the transition cascades but also to suppress the background drastically. PARIS has high detection efficiency in the 6$-$10 MeV range. The energy resolution and detection efficiency of PARIS clusters are 112 keV and 0.30(2)$\%$, respectively, at 7.18 MeV, under mode 3; signals only come from LaBr$_3$(Ce) crystals and are vetoed by outer-layer NaI detectors, but the energy is added back within 27 phoswich detectors. The detection efficiency of HPGe detectors including one clover and one coaxial detector is 3.134(5)$\%$ at 1085.836 keV. For more details of detection efficiencies of HPGe and PARIS detectors in different data sorting modes, e.g., with anticoincidence and without anticoincidence using the NaI outer layer, we refer to Figs. 3.10 and 3.16 in Ref. \cite{ren2022thesis}. Furthermore, one is able to reject pileup events and significantly suppress Compton-scattered escaping events using PARIS clusters, covering angles 0$^{\circ}-$56.3$^{\circ}$ when a photon hits the center of the first crystal. Certainly, one should count the detection efficiency of the outer-layer NaI as well. Thanks to the larger volume of the outer-layer NaI crystal of 2 in.$\times$2 in.$\times$6 in. compared to 2 in.$\times$2 in.$\times$2 in. for the front LaBr$_3$(Ce) crystal of a PARIS detection unit, we have higher detection efficiency in the outer layer to perform anti coincidence analysis. According to the Klein-Nishina formula, for a $\gamma$ ray with energy larger than 2 MeV, the scattering cross section for an angle larger than 56.3$^{\circ}$ is rather small. Furthermore, the scattered photons to large angles are mainly in the 0$-$2 MeV energy range. Thus, the higher-energy part of the $\gamma$ spectrum is background free. One can directly extract information avoiding the use of Monte Carlo simulations, e.g., by calculating the dominant background of the summing-pileup spectrum. Notably, summing-pileup differs from pileup. The latter refers to electronic pulse pileup (pulses overlap) for a single crystal due to more than one event arriving within the analog-to-digital converter (ADC) gate \cite{CANOOTT1999488} while the former refers to the summing of signals from different detector modules \cite{PhysRevLett.115.062502,PhysRevC.95.024320}. The normalization of the simulated summing-pileup spectrum is fixed using the event rate and the ADC gate length; see more details about normalization factor calculations and the comparison with measurements in Ref. \cite{CANOOTT1999488}. The detectors were energy calibrated up to 9 MeV using sources including $^{152}$Eu, AmBe, and $^{58}$Ni(n$_\textnormal{th}$,$\gamma$) with energy of 8999.267(15) keV. Eventually, the detection set up covered the whole $\it{Q}_{\beta}$ window. Therefore, a detailed spectroscopy of low- and high-energy states has been achieved.

Data were acquired in a triggerless mode. The decay level schemes were constructed using the $\beta$-$\gamma$-$\gamma$ coincidence technique. All levels are built with HPGe $\gamma$-$\gamma$ information except for the 7840 and 7996 keV states which are obtained with a combination of HPGe and PARIS $\gamma$-ray data. The decay level schemes were experimentally separated using the "$\it{X}$" method which is based on the measurement of the apparent half-life of each individual $\gamma$ line. $\it{X}$ is defined as the fractional direct $\beta$ feeding from $^{80m}$Ga. For a given state, the apparent half-life or decay constant ($\lambda_A$) is the result of several contributions: the fractional direct $\beta$ feeding from $^{80g}$Ga [Br$_{{\beta}F}$(1-$\it{X}$)], the fractional direct $\beta$ feeding from $^{80m}$Ga (Br$_{{\beta}F}\it{X}$), the half-life of $^{80m}$Ga ($\it{T}_{1/2}^S$), the half-life of $^{80g}$Ga ($\it{T}_{1/2}^L$), the branching ratio of $\gamma$ feeding (Br$_{{\gamma}F}$), and the weighted half-life of $\gamma$ rays involved in this $\gamma$ feeding ($\it{T}_{1/2}^{{\gamma}F}$). Equation ($\ref{X1}$) relates these parameters \cite{verney2013structure,ren2022thesis}:

\begin{equation} \label{X1} 	
\lambda_A = \frac{\textnormal{Br}_{{\beta}F}}{\textnormal{Br}_{{\beta}F} + \textnormal{Br}_{{\gamma}F}}[X\lambda_S + (1-X)\lambda_L] + \frac{\textnormal{Br}_{{\gamma}F}}{\textnormal{Br}_{{\beta}F} + \textnormal{Br}_{{\gamma}F}}\lambda_{{\gamma}F}
\end{equation}

where $\lambda$ = $\ln$2/$\it{T}_{1/2}$, and $\it{T}_{1/2}^S$ and $\it{T}_{1/2}^L$ are the half-lives of $^{80m}$Ga and $^{80g}$Ga, respectively. $\it{T}_{1/2}^{{\gamma}F}$ was taken as the weighted average of all the $\gamma$ rays of indirect feedings. Br$_{{\beta}F}$ + Br$_{{\gamma}F}$ = 1 for a given state. Then, one derives the equation of $\it{X}$ as

\begin{equation} \label{X2} 	
X = \frac{1}{R}\frac{1/T_{1/2}^{{\gamma}F} - 1/T_{1/2}^A}{1/T_{1/2}^L - 1/T_{1/2}^S} + \frac{1/T_{1/2}^L - 1/T_{1/2}^{{\gamma}F}}{1/T_{1/2}^L - 1/T_{1/2}^S} 
\end{equation}

where $\it{R}$ = Br$_{{\beta}F}$/(Br$_{{\beta}F}$ + Br$_{{\gamma}F}$) is the proportion of the direct $\beta$-feeding contribution in the total direct and indirect feedings of one given state. It can be calculated using the balance method, i.e., the difference between the $\gamma$-ray counts of deexcitations and $\gamma$-feedings:

\begin{equation} \label{X5} 	
	R = \frac{C_{all{\hspace{0.035in}}{\gamma}{\hspace{0.035in}}of{\hspace{0.035in}}deexcitations}-C_{{\gamma}{\hspace{0.035in}}from{\hspace{0.035in}}{\gamma}{\hspace{0.035in}}feedings}}{C_{all{\hspace{0.035in}}{\gamma}{\hspace{0.035in}}of{\hspace{0.035in}}deexcitations}}
\end{equation}

\begin{figure}
    \centering
    \includegraphics[width=1.0\columnwidth]{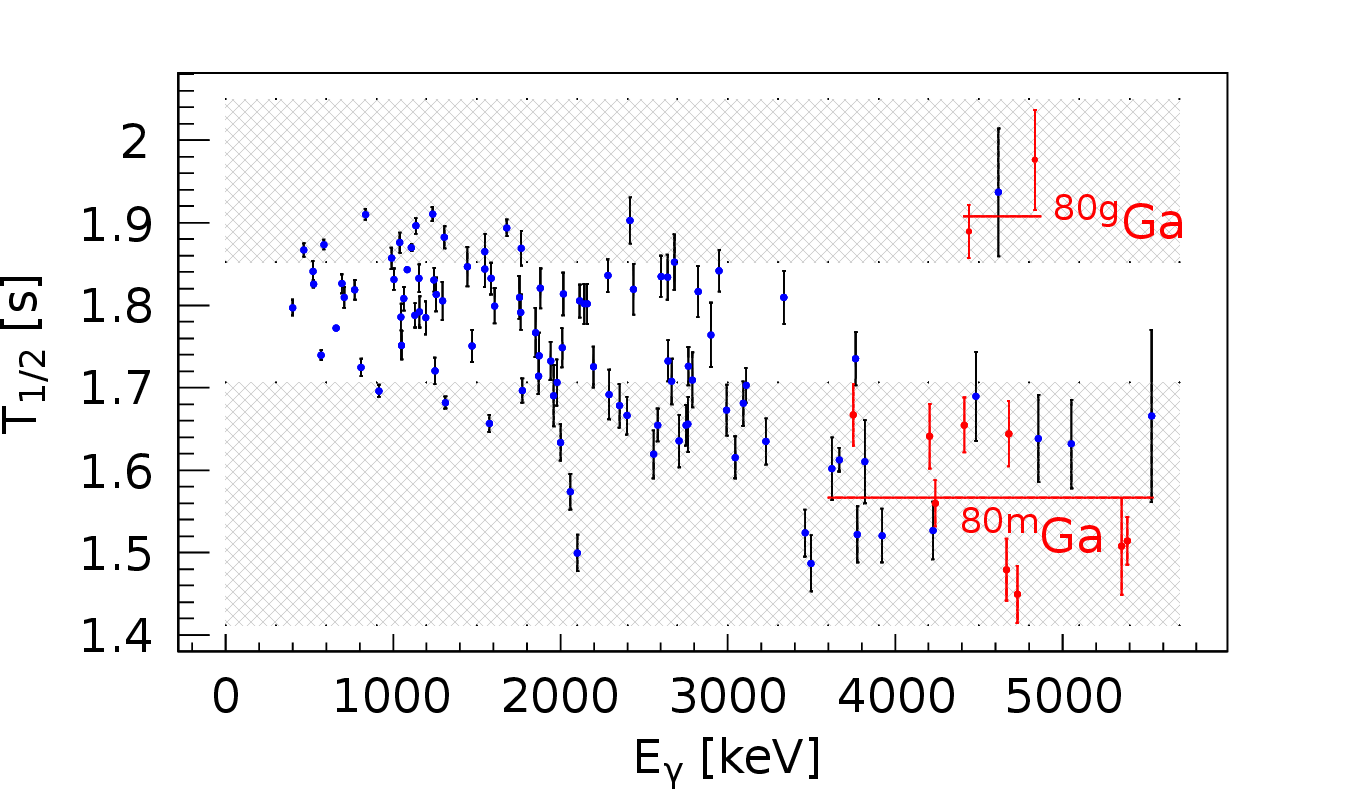}
    \caption{Measured half-lives of all individual $\gamma$-line activities associated with the decay of the mixed $^{80g+m}$Ga source. Red points show data used to extract half-lives of $^{80g}$Ga and $^{80m}$Ga. The red lines are the fittings of red points.}
    \label{fig00}
\end{figure} 

$\it{T}_{1/2}^A$ is the measured apparent half-life of one given state. Note that the precision on the extracted $\it{X}$ values relies primarily on the available statistics, the capability to detect all indirect feedings including the ones from high energy region. Since the spin-parity values of the two precursors are 3$^-$ and 6$^-$, and considering allowed and first-forbidden non unique (ffnu) transitions, there are no overlaps between states directly $\beta$ fed by the two isomers. This is because they populate (2,3,4)$^{-,+}$ states and (5,6,7)$^{-,+}$ states, respectively. An overlap would be only possible when one of the two isomers decays via first-forbidden unique (ffu) transition or higher-order transition such as second-forbidden non unique/unique transitions. Given the much smaller probability of ffu compared to allowed and ffnu transitions, we assigned crossed $\beta$ feedings to a state using the narrow window of $\it{X}$ value, 0.4$-$0.6. Levels with $\it{X}$ = 0 are considered perfect candidates for the decay scheme of $^{80g}$Ga and $\it{X}$ = 1 for the decay scheme of $^{80m}$Ga. In addition to ideal situations, the actual actions taken are (1) $\it{x}$ < 0.4 are from $^{80g}$Ga decay; (2) $\it{x}$ > 0.6 are from $^{80m}$Ga decay; (3) 0.4 < $\it{x}$ < 0.6, 50$\%$ from $^{80g}$Ga and 50$\%$ from $^{80m}$Ga. Finally, only ten states out of a total of 77 populated states were assigned to $\beta$ feedings from both $^{80g}$Ga and $^{80m}$Ga simultaneously. They are the levels with the energies of 1972.2, 3423.4, 3515.5, 3610.7, 3721.1, 3752.2, 4029.3, 4408.2, 4462.5, and 5806.3 keV. The summed $\it{I}_{\beta}$ of these ten states are 8.1(6)$\%$ and 13.9(8)$\%$ in decay level schemes of $^{80g}$Ga and $^{80m}$Ga, respectively. It is worth pointing out that having the intensity distributed in both level schemes for some $\gamma$ rays and levels can introduce a systematic error in the feedings. Further checks of these ten states with high statistics deserve being pursued in future experiments.

\begin{table}[htbp!] 
\centering
\caption{$\gamma$-transitions in the decay of $^{80g+m}$Ga to $^{80}$Ge. The uncertainties are given in nuclear data sheets format.}  \vspace{0.5mm}
\label{tab3}
\renewcommand\arraystretch{0.9}
\begin{tabular}{llll}
$E_{\gamma}$ (keV) \hspace{0.3 in} & $\it{I}^{\textnormal{rel}}_{\gamma}$ \hspace{0.3 in} & $\it{E}^{i}_\textnormal{level}$ (keV) \hspace{0.3 in} & E$^{f}_\textnormal{level}$ (keV) \\
\hline
\hline
398.6 3		& 0.35 4	&1972.2 4	&1574.1 4 \\
466.7 3		&1.59 5		&3445.3	6	&2978.5 5 \\
520.0 3		&1.35 5		&3498.6 6	&2978.5 5 \\
523.1 3		&14.41 40	&2266.0 5	&1742.9	4 \\
571.1 3		&5.90 17	&3423.4 6	&2852.3 5 \\
586.2 3		&8.48 24	&2852.3 5	&2266.0 5 \\
659.2 3		&100.0 28	&659.2 3	&0.0 0 \\
692.6 3		&0.61 3		&2266.0 5	&1574.1 4 \\
707.9 3		&0.25 3		&3686.9 6	&2978.5	5 \\
771.6 3		&0.54 3		&3037.2	5	&2266.0 5 \\
809.1 3		&0.77 5		&4324.2 6	&3515.5 6 \\
834.6 3		&7.46 22	&3686.9 6	&2852.3	5 \\
915.1 3		&4.22 13	&1574.1 4	&659.2 3 \\
989.8 3		&1.61 6		&4413.1	7	&3423.4 6 \\
1005.2 3	&1.35 6		&4993.5 7	&3988.3 6 \\
1040.9 3	&2.15 7		&5573.7 7	&4532.8	6 \\
1047.7 4	&0.18 5		&4026.2	6	&2978.5	5 \\
1050.8 4	&0.05 4		&4029.3 7	&2978.5	5 \\
1064.9 3	&0.98 5		&3037.2	5	&1972.2	4 \\
1083.6 3	&68.3 20	&1742.9 4	&659.2 3 \\
1109.5 3	&28.43 85	&2852.3	5	&1742.9	4 \\
1130.7 3	&1.31 5		&3396.8 6	&2266.0 5 \\
1136.0 3	&5.60 17	&3988.3 6	&2852.3	5 \\
1154.7 3	&0.87 7		&5567.8	7	&4413.7	7 \\
1157.5 4	&0.32 9		&3424.0 6	&2852.3 5 \\
1194.8 3	&0.24 3		&4173.3	6	&2978.5	5 \\
1235.7 3	&7.36 23	&2978.5	5	&1742.9	4 \\
1244.8 3	&1.17 5		&5233.1 7	&3988.3 6 \\
1249.3 3	&0.52 3		&3515.5 6	&2266.0 5 \\
1257.6 3	&0.40 3		&4944.6 7	&3686.9 6 \\
1294.2 3	&1.04 4		&3037.2 5	&1742.9	4 \\
1306.7 3	&2.92 9		&4993.5 7	&3686.9 6 \\
1312.9 3	&6.91 22	&1972.2 4	&659.2 3 \\
1444.3 4	&0.17 3		&4422.6	6	&2978.5	5 \\
1471.8 3	&0.55 4		&4324.2 6	&2852.3 5 \\
1547.2 4	&0.50 3		&5233.1 7	&3686.9	6 \\
1548.5 5	&0.50 3		&4993.5 7	&3445.3 6 \\
1573.6 3	&3.41 8		&1574.1 4	&0.0 0 \\
1585.3 3	&1.06 4		&5573.7 7	&3988.3	5 \\
1606.7 4	&0.24 7		&2266.0 5	&659.2 3 \\
1680.5 3 	&6.90 15	&4532.8	6	&2852.3 5 \\
1753.5 4	&0.30 3		&4732.1	6	&2978.5 5 \\
1763.5 3	&0.40 24	&4615.8 6	&2852.3 5 \\
1766.7 5	&0.64 25	&5453.6	8	&3686.9 6 \\
1772.6 4	&1.49 4		&3515.5 6	&1742.9	4 \\
1850.2 5	&1.00 4		&3424.0 6	&1574.1 4 \\
1867.8 7	&0.27 22	&3610.7 8	&1742.9	4 \\
1881.0 4	&0.32 3		&5567.8	7	&3686.9 6 \\
1941.7 4	&0.47 4		&3515.5 6	&1574.1 4 \\
1978.3 5	&0.13 2		&3721.1	6	&1742.9	4 \\
1999.7 4	&0.39 3		&4851.8	6	&2852.3	5 \\
2009.3 4	&0.50 5		&3752.2 6	&1742.9	4 \\
2017.0 4	&0.38 5		&5703.8	7	&3686.9 6 \\
2058.9 4	&0.15 2		&4324.2	6	&2266.0 5 \\
2101.9 4	&0.22 3		&4073.3 7	&1972.2	4 \\
2115.0 4	&1.33 4		&5801.1	7	&3686.9 6 \\
\hline

\end{tabular}
\end{table}

\begin{table}[htbp!] 
\ContinuedFloat
\centering
\caption{($\it{Continued}$)}  \vspace{0.5mm}
\label{tab3}
\renewcommand\arraystretch{0.9}
\begin{tabular}{llll}
$E_{\gamma}$ (keV) \hspace{0.3 in} & $\it{I}^{\textnormal{rel}}_{\gamma}$ \hspace{0.3 in} & $\it{E}^{i}_\textnormal{level}$ (keV) \hspace{0.3 in} & E$^{f}_\textnormal{level}$ (keV) \\
\hline
\hline
2141.0 4	&0.74 6		&4993.5 7	&2852.3	5 \\
2160.9 3	&0.54 10	&6187.1	7	&4026.2 6 \\
2196.5 4	&0.55 5		&4462.5	6	&2266.0 5 \\
2283.3 4	&1.73 5		&4026.2 6	&1742.9	4 \\
2290.9 5	&0.13 2		&5806.3	7	&3515.5 6 \\
2351.7 4	&0.30 3		&4324.2	6	&1972.2	4 \\
2396.5 4	&0.53 4		&4139.4 6	&1742.9	4 \\
2418.9 4	&0.44 3		&6407.2 7	&3988.3 6 \\
2436.5 5	&0.43 4		&4179.3	6	&1742.9	4 \\
2555.1 5	&0.27 3		&3214.4 5	&659.2 3 \\
2581.1 4	&1.04 4		&4324.2	6	&1742.9	4 \\
2600.2 5	&1.08 4		&5452.2 7	&2852.3	5 \\
2638.4 5	&0.53 4		&5490.7 7	&2852.3	5 \\
2642.4 5	&1.10 5		&3301.6 5	&659.2 3 \\ 
2665.4 5	&0.54 3		&4408.2	6	&1742.9	4 \\
2679.6 5	&0.26 3		&4422.6	6	&1742.9	4 \\
2709.9 5	&0.12 3		&4682.1 7	&1972.2 4 \\
2750.4 5	&0.66 4		&4324.2	6	&1574.1 4 \\
2762.3 5	&0.11 2		&6185.7 8	&3423.4 6 \\
2764.5 5	&0.82 4		&3424.0 6	&659.2 3 \\
2789.3 6	&0.07 2		&4532.8	6	&1742.9	4 \\
2821.9 5	&0.56 3		&5801.1 7	&2978.5	5 \\
2948.2 5	&1.35 4		&5801.1	7	&2852.3	5 \\
2993.2 5	&0.41 3		&4736.1 6	&1742.9	4 \\
3044.1 5	&0.43 4		&3703.4 6	&659.2 3 \\
3091.1 5	&0.42 3		&3750.3 6	&659.2 3 \\
3108.9 5	&1.17 4		&4851.8	6	&1742.9	4 \\
3228.0 5	&0.40 3		&3887.3 6	&659.2 3 \\
3336.0 6	&0.48 8		&6187.1	7	&2852.3	5 \\
3461.8 6	&0.09 3		&5035.9	7	&1742.9	4 \\
3498.1 6	&0.22 4		&5072.1	7	&1574.1	4 \\
3621.7 6	&0.18 3		&5364.5 7	&1742.9 4 \\
3664.6 6	&3.39 10	&4324.2	6	&659.2 3 \\
3750.5 6	&0.26 4		&5324.5 8	&1574.1 4 \\
3764.3 6	&0.49 6		&5338.1	7	&1574.1 4 \\
3774.4 7	&0.22 5		&5517.2 8	&1742.9 4 \\
3818.6 6	&0.64 6		&4477.8 7	&659.2 3 \\
3920.4 6	&0.49 4		&4579.6	7	&659.2 3 \\
4207.8 7	&0.25 4		&6473.8	8	&2266.0 5 \\
4225.1 8	&0.21 5		&4884.4	8	&659.2 3 \\
4238.8 7	&0.57 5		&6211.0	8	&1972.2	4 \\
4412.8 7	&0.68 4		&5072.1	7	&659.2 3 \\
4443.6 7	&1.46 6		&6187.1	7	&1742.9	4 \\
4482.9 7	&0.30 2		&6057.0 8	&1574.1 4 \\
4617.9 9	&0.09 3		&5277.1	9	&659.2	3 \\
4665.3 8	&0.50 5		&5324.5 8	&659.2	3 \\
4678.5 7	&0.51 4		&5338.1	7	&659.2	3 \\
4729.6 7	&0.33 3		&6472.5	8	&1742.9	4 \\
4835.7 8	&0.33 4		&6578.5 9	&1742.9	4 \\
4856.1 7	&0.27 3		&6599.6	8	&1742.9	4 \\		
5053.0 8	&0.34 3		&6627.0 9	&1574.1 4 \\
5353.7 8	&0.28 3		&6013.0	9	&659.2 3 \\
5387.4 8	&1.29 5		&6046.6	8	&659.2 3 \\
5531.5 9	&0.17 3		&6190.7	9	&659.2 3 \\
7181 71		&0.48 3		&7840 71	&659.2 3 \\
7337 72		&0.14 1		&7996 72	&659.2 3 \\
\hline
\hline

\end{tabular}
\end{table}

\begin{figure}[!htb]
\includegraphics[width=1.0\columnwidth]{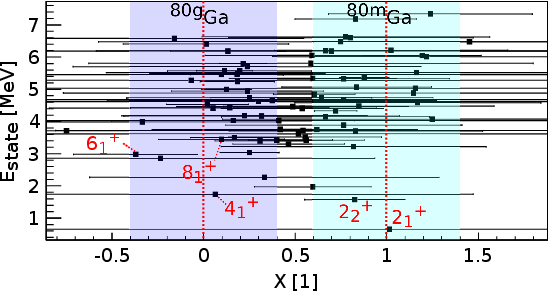}
\caption{$\it{X}$ values of $\beta$-populated levels. Well-identified states 2$_1^+$, 2$_2^+$, 4$_1^+$, 6$_1^+$, and 8$_1^+$ are marked with red numbers.}
\label{fig01}
\end{figure}

Figure \ref{fig00} presents the half-lives of individual $\gamma$-line activities as functions of their energies. In this figure, one observes a clear splitting in the high-energy region and the scattering between the half-lives of the ground state and the isomer in the lower-energy region. In the present work, a total of 112 characteristic half-lives of activities of $\beta$-delayed $\gamma$ rays could be determined. They are tabulated in Table \ref{tab3}. Among them 70 were assigned to the decay of $^{80g}$Ga and 65 to the decay of $^{80m}$Ga. Twenty-three $\gamma$-rays are emitted by both isomers. These are the lines at 398.6, 523.1, 571.1, 586.2, 659.2, 692.6, 915.1, 1050.8, 1083.6, 1109.5, 1236.7, 1249.3, 1312.9, 1573.6, 1606.7, 1772.6, 1867.8, 1941.7, 1978.3, 2009.3, 2196.5, 2290.9, and 2665.4 keV. The red points represent the selected $\gamma$ lines for extracting the half-life of $^{80g}$Ga, i.e., the 4443.6(7) keV depopulating the 6187.1(7) keV state and the 4835.7(7) keV from 6578.5(9) keV state, and the half-life of $^{80m}$Ga, i.e. the 3750.5(6), 4207.8(7), 4238.8(7), 4412.8(7), 4665.3(8), 4678.5(7), 4729.6(7), 5387.4(8), and 5353.7(8) keV $\gamma$ rays, which deexcite from high-lying states as well. For the criterion of selecting these $\gamma$ lines, we refer to the first section of Chap. 4 of Ref. \cite{ren2022thesis}. Finally, the extracted half-lives of $^{80g}$Ga and $^{80m}$Ga are 1.91(3) and 1.57(1) s, respectively.

Figure \ref{fig01} shows the calculated $\it{X}$ values of states $\beta$ populated by $^{80g+m}$Ga, which were used to attribute each state to one of the two decay schemes.

After calculating the $\it{X}$ values, we applied two other methods as additional verification criteria for our identification. The first one is to compare the gated $\gamma$ spectra under different conditions on the $\beta$-activity curve. This profits from the difference in half-lives between the ground state and isomer. For example, one can put a gate on 5.5$-$8 s (period 1), after beam collection, to obtain a gated spectrum. Next, the same operation can be taken but gated on 8$-$10.5 s (period 2) to get another gated spectrum. Then, one can immediately observe some surviving peaks in period 2 being relatively weaker than others. This demonstrates that the precursor populating these $\beta$-delayed $\gamma$ rays has a shorter half-life than the other one. This would be $^{80m}$Ga.

\begin{figure*}[!htb]
\centering
\includegraphics[width=1.0\textwidth]{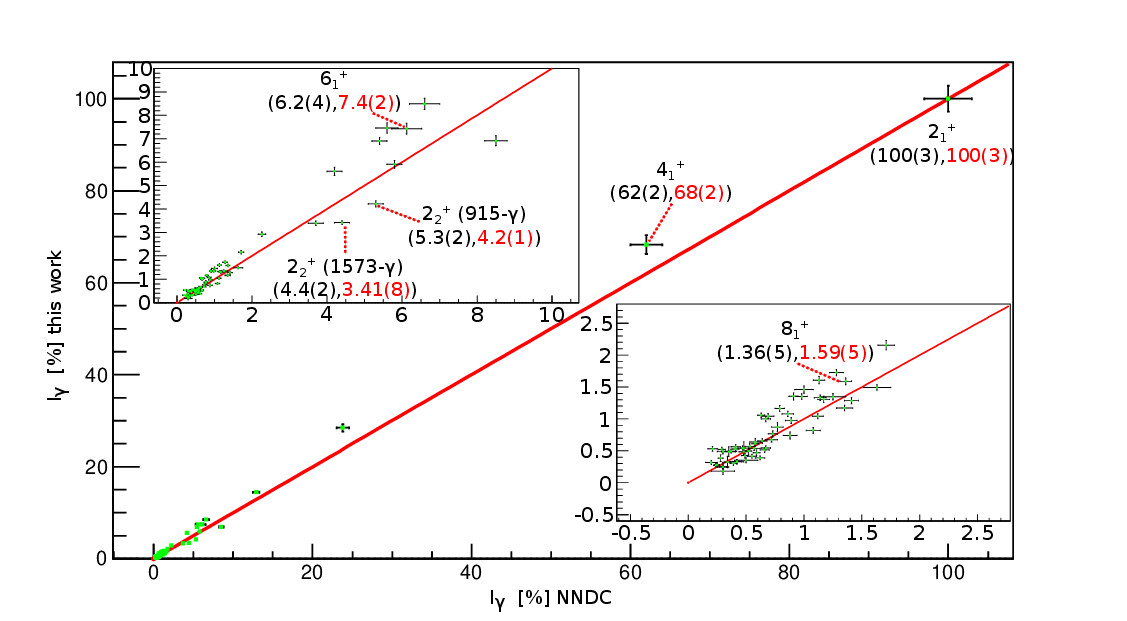}
\caption{Relative $\gamma$ intensities of transitions in $^{80}$Ge observed in this work from photo fission of UC$_x$ (vertical axis) as a function of relative $\gamma$ intensities from the evaluated values \cite{ENSDF} from thermal neutron-induced fission of $^{235}$U \cite{hoff1981properties} (horizontal axis). The two insets are zoom-in figures.}
\label{fig02}
\end{figure*}

The second one is to compare the relative $\gamma$-ray intensities in different fissioning systems as they produce different isomeric ratios. For example, through this comparison, one finds that the $^{80m}$Ga ratio is lower in photo fission UC$_x$($\gamma$,n$_f$,$\it{f}$) than in thermal neutron-induced fission $^{235}$U(n$_\textnormal{th}$,$\it{f}$). Therefore, if a $\gamma$-ray relative intensity is lower in photo fission, one can reach the conclusion that the related state emitting this $\gamma$ ray was $\beta$ fed by $^{80m}$Ga. Fig. \ref{fig02} presents the relative intensities of the $\beta$-delayed $\gamma$ rays from this work compared with thermal neutron data $\cite{hoff1981properties}$. One observes that there is a distribution relative to the "$\it{X}$ = $\it{Y}$" line. Therefore, $\gamma$ rays located above this line can be generally assigned to $^{80g}$Ga $\beta$ feedings and those below this line are assigned to $^{80m}$Ga $\beta$ feedings. $\gamma$ lines from states with well-identified spin-parity are marked. No inconsistencies were found between these three methods in precursor identification.

\section{III. Results and discussion}

Formula ($\ref{beta-decay}$) was used to extract $\it{B}$(GT) from observed log $\it{ft}$ values. Note that for $^{80}$Ga the Fermi terms are zero as the isobaric analogue state is located above $\it{Q}_{\beta}$:

\begin{multline} \label{beta-decay}
{\fontsize{9}{2}\selectfont
\begin{aligned}
t_{1/2}^{-1}=&\frac{1}{K}fg^2_A\frac{|<f||\sum\limits_{i=1}^{A}\sigma{\tau}^-||i>|^2}{2J_i+1} \\
	=&\frac{1}{K}fg^2_A|M^{eff}_{GT}|^2 \\
	=&\frac{1}{K}fg^2_AB(GT) 
\end{aligned}
}
\end{multline}

Where $\it{K}$/($\hbar$c)$^6$=2${\pi}^3\hbar$ln2/($\it{m}_e$c$^2$)$^5$=8120.27648(26)$\times$10$^{-10}$ \cite{hardy2020superallowed}, $\it{f}$ is the Fermi integral function, $\it{M}_\textnormal{F/GT}^\textnormal{eff}$ is the effective nuclear matrix element and $\it{g}_A$ is the axial-vector coupling constant in units of Fermi constant $\it{G}_F$, $\it{g}_A$/$\it{g}_V$=-1.2756(13) \cite{Workman2022}.

\begin{figure*}[!htb]
\centering
\includegraphics[width=1.0\textwidth]{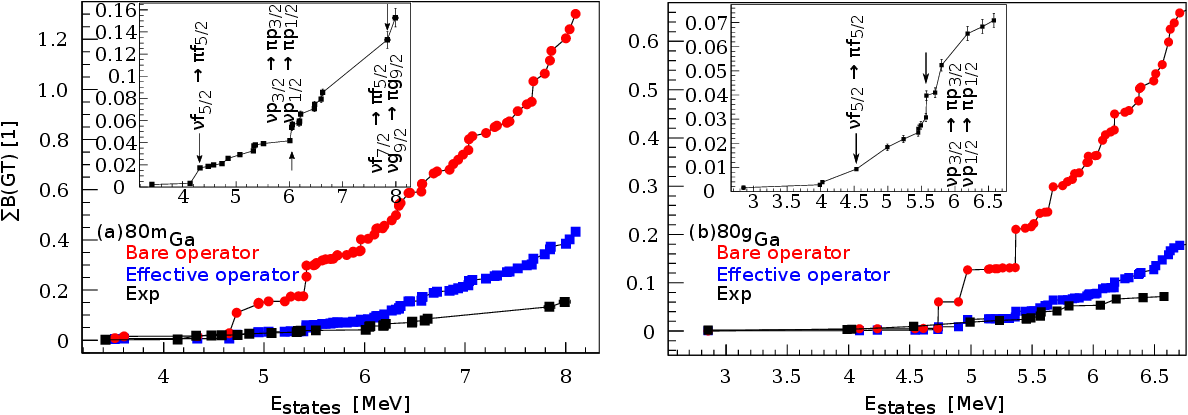}
\caption{Experimentally measured cumulative $\it{B}$(GT) (black squares) with statistical uncertainties of $^{80m}$Ga (a) and $^{80g}$Ga (b) versus excitation energies of the final states in $^{80}$Ge. The experimental results obtained in this work are compared to the realistic shell-model calculations performed with the bare GT operator (red dots) and with the effective GT operator (blue squares). The insets are zoom-in figures of the black curves.}
\label{fig1}
\end{figure*}

Figure \ref{fig1}(a) and Figure \ref{fig1}(b) present the cumulative $\it{B}$(GT) up to $\it{S}_n$ (8.08 MeV). The related states GT-populated by $^{80m}$Ga and $^{80g}$Ga are listed in Table \ref{tab1}. The total $\it{B}$(GT) within this energy window is measured to be 0.152(8) for $^{80m}$Ga and 0.072(3) for $^{80g}$Ga. The uncertainties on $\it{B}$(GT) values originate from the uncertainties on $\it{Q}_{\beta}$ value, half-lives of precursors, $\it{I}_{\beta}$ and excitation energies of states. For cumulated results, the method of propagation of uncertainty was used. One can find that two newly measured states located just below $\it{S}_n$ have much larger $\it{B}$(GT) values than others although the $\it{I}_{\beta}$'s are not very large. It is because their Fermi integral phase spaces ($\it{f}$) are smaller than for low-lying states.

\begin{table}[htbp] 
\centering
\caption{Excited states of $^{80}$Ge GT populated by $^{80m}$Ga and $^{80g}$Ga, before and after the horizontal line, respectively. The $\it{B}$(GT) values are presented in Fig. \ref{fig1}. The uncertainties are given in nuclear data sheets format.}  \vspace{0.5mm}
\label{tab1}
\renewcommand\arraystretch{0.7}
\begin{tabular}{llllll}
$E_\textnormal{state}$ (keV) \hspace{0.2 in}	& $\it{I}_{\beta}$ \hspace{0.2 in}	& log $\it{ft}$ \hspace{0.2 in}		& $\it{J}^{\pi}$ \hspace{0.2 in}	 & "$\it{X}$"	\\
\hline
\hline
3423.4 6 \hspace{0.2 in}		& <4.5 5 \hspace{0.2 in}		& >6.23 5 \hspace{0.2 in}		& (3)$^-$ \hspace{0.2 in}	& 0.56 32	\\
4139.4 6 		& 1.1 1 			& 6.62 5 		& (3,4)$^{-}$ 	& 0.67 57	\\
4324.2 6		& 14.6 16		& 5.43 5		& (3$^{-}$)	& 0.72 12	\\
4477.8 7		& 1.4 2			& 6.40 5		& (2,3)$^{-}$	& 0.85 74 	\\
4579.6 7		& 1.0 1			& 6.52 5		& (2,3)$^{-}$	& 1.17 67	\\
4736.1 6		& 0.9 1			& 6.51 5		& (3,4)$^{-}$	& 0.65 61	\\
4851.8 6		& 3.3 4			& 5.90 6		& (3,4)$^{-}$	& 0.60 22	\\
5072.1 7		& 1.9 2			& 6.06 5		& (2,3)$^{-}$	& 0.84 41	\\
5324.5 8		& 1.6 2			& 6.04 5		& (2,3)$^{-}$	& 0.76 58 	\\
5338.1 7 		& 2.1 2			& 5.93 6		& (2,3)$^{-}$	& 0.60 46	\\
5364.5 7		& 0.39 5		& 6.64 6		& (3,4)$^{-}$	& 0.88 14	\\
5517.2 8		& 0.46 6		& 6.50 6		& (3,4)$^{-}$	& 1.2 17	\\
6013.0 9		& 0.60 7		& 6.18 5		& (2,3)$^{-}$	& 1.22 84	\\
6046.6 8		& 2.7 3			& 5.51 5		& (2,3)$^{-}$	& 1.19 32	\\
6057.0 8		& 0.64 7		& 6.15 5		& (2,3)$^{-}$	& 0.59 58 	\\
6185.7 8		& 0.24 3 		& 6.50 6		& (3,4)$^{-}$	& 0.7 16	\\
6190.7 9		& 0.36 5		& 6.32 6		& (2,3)$^{-}$	& 0.7 13	\\
6211.0 8		& 1.2 1			& 5.79 5		& (3,4)$^{-}$	& 1.02 64	\\
6472.5 8		& 0.71 8		& 5.89 5		& (3,4)$^{-}$	& 1.45 85	\\
6473.8 8		& 0.53 7		& 6.01 6		& (3,4)$^{-}$	& 0.7 13	\\
6599.6 8		& 0.57 7		& 5.92 6    	& (3,4)$^{-}$	& 0.80 67	\\
6627.0 9		& 0.72 8		& 5.80 5	    & (2,3)$^{-}$	& 0.78 79	\\
7840 71 		& 1.0 1			& 4.92 6		& (2,3)$^{-}$	& 0.83 33	\\
7996 72			& 0.31 3		& 5.31 6		& (2,3)$^{-}$	& 1.24 56	\\
\hline
2852.3 5		& <5.5 9		& >6.38 8		& (5$^{-}$)	& $-$0.2 12	\\
3988.3 6		& <1.9 3		& >6.50 6		& (5,6)$^{-}$	& $-$0.33 74	\\
4026.2 6		& <1.7 2		& >6.55 6		& (5$^{-}$)	& 0.16 66	\\
4532.8 6		& 5.9 7			& 5.84 5		& (5,6)$^{-}$	& 0.02 22		\\
4993.5 7 		& 6.8 7			& 5.62 5		& (6,7)$^{-}$	& 0.12 14		\\
5233.1 7		& 2.1 2			& 6.05 5		& (5,6,7)$^{-}$	& 0.18 20	\\
5452.2 7		& 1.3 2			& 6.15 5		& (5,6,7)$^{-}$	& 0.18 25	\\
5453.6 8		& 0.8 2			& 6.4 1			& (5,6,7)$^{-}$	& 0.1 26	\\
5490.7 7		& 0.66 8		& 6.44 6		& (5,6,7)$^{-}$	& 0.19 48	\\
5567.8 7		& 1.5 2 		& 6.06 5		& (5,6,7)$^{-}$	& 0.20 39	\\
5573.7 7		& 4.0 4			& 5.62 5		& (5,6,7)$^{-}$	& 0.11 15	\\
5703.8 7		& 0.47 7		& 6.50 7		& (5,6,7)$^{-}$	& 0.24 95	\\
5801.1 7		& 4.0 4			& 5.52 5		& (5,6,7)$^{-}$	& 0.22 13	\\
6187.1 7		& 3.1 4			& 5.46 5	    & (5$^{-}$)		& 0.13 32	\\
6407.2 7		& 0.54 6		& 6.12 5		& (5,6,7)$^{-}$	& 0.01 45	\\
6578.5 9		& 0.40 5		& 6.16 6		& (5)$^{-}$		& $-$0.16 71	\\

\end{tabular}
\end{table}

Furthermore, from experimental data, one observes clear jump-structure in $\Sigma\it{B}$(GT) functions of both precursors. These jumps can be interpreted as doorways of single-particle GT $\beta$-transitions, based on a single-particle picture \cite{grawe2004shell}, which are dominated by (1) $\nu\it{f}_{5/2}$ $\rightarrow$ $\pi\it{f}_{5/2}$ (first jump); (2) $\nu\it{p}_{3/2}$ $\rightarrow$ $\pi\it{p}_{3/2}$ or/and $\nu\it{p}_{1/2}$ $\rightarrow$ $\pi\it{p}_{1/2}$ (second one); (3) $\nu\it{f}_{7/2}$ $\rightarrow$ $\pi\it{f}_{5/2}$ or/and $\nu\it{g}_{9/2}$ $\rightarrow$ $\pi\it{g}_{9/2}$ (third one), respectively.

The experimental data are compared to theoretical calculations performed within a realistic shell-model approach. In particular, we have considered a $^{56}$Ni core with the model space spanned by 0$\it{f}_{5/2}$, 1$\it{p}_{3/2}$, 1$\it{p}_{1/2}$ and 0$g_{9/2}$ for both protons and neutrons. The two-body matrix elements of the effective Hamiltonian have been derived, within the framework of many-body perturbation theory, starting from the CD-Bonn potential \cite{machleidt2001high} renormalized by way of the V$_\textnormal{low-k}$ approach \cite{BONGER2001432} with the addition of the Coulomb term for the proton-proton interaction. More precisely, the $\hat{Q}$-box folded-diagram approach was employed \cite{coraggio2012effective} including one- and two-body diagrams up to the third order in the interaction in the perturbative diagrammatic expansion of the $\hat{Q}$ box. As regards the GT operator, it is well known that the diagonalization of the effective Hamiltonian does not produce the true nuclear wave functions, but their projections onto the selected model space. As a consequence, any bare decay operator should be renormalized by taking into account the neglected degrees of freedom. For this purpose, we use the Suzuki-Okamoto formalism \cite{suzuki1995effective}. This allows a derivation of the decay operator consistent with the effective Hamiltonian. Consequently, the effective charges are state dependent as can be seen from Table XVII of Ref. \cite{coraggio2019renormalization}, where the GT matrix elements of the effective GT operator are reported together with the corresponding quenching factors. The effective Hamiltonian and GT operator so derived have already been used in Ref. \cite{coraggio2019renormalization} to study the GT and two-neutrino double-$\beta$ decay matrix elements of $^{76}$Ge and $^{82}$Se. Finally, we stress that our calculation is fully microscopical. In fact, we do not resort to any empirical quenching factor for the axial coupling constant $\it{g}_A$.

The microscopical nature of our calculations, and the large number of valence nucleons, may have as a consequence a not-perfect reproduction of the energy levels. Therefore, in the comparison between theoretical and experimental $\Sigma\it{B}$(GT) the energy of the lowest state has been shifted by 0.4899 and 0.4811 MeV, respectively, for $^{80g}$Ga and $^{80m}$Ga to reproduce the energy of the corresponding experimental levels. As can be seen in Figs. \ref{fig1}(a) and \ref{fig1}(b), the theoretical calculations produce similar jump structures in the running sums.

The effect of the renormalization is very evident in both cases, as it was for $^{76}$Ge and $^{82}$Se GT strength \cite{coraggio2019renormalization}. In cases of $^{80m+g}$Ga, up to 6 MeV, we find that the theoretical strength is quenched by a factor of $\approx$0.28 for the isomeric state and by a factor of $\approx$0.27 for the ground state, respectively, under renormalization.

This strong renormalization is not surprising. In fact, while a phenomenological quenching factor of 0.744(15) for $\it{g}_A$ is usually needed to reproduce the experimental data in the region of $\it{fp}$-shell nuclei \cite{Pinedo1996}, a higher renormalization is required in heavier-mass-region nuclei, as shown in a previous study \cite{coraggio2019renormalization}. This is because, for the spin- and spin-isospin-dependent operators like \textbf{$\Sigma\sigma{\tau}^-$}, a configuration with more than one valence nucleon plays a more important role in nuclei located in medium- and heavy-mass regions than those in the light-mass region. This conclusion is consistent with observation of 2$_1^+$-based pygmy dipole resonance that evidences collectivity in the daughter nucleus $^{80}$Ge \cite{ren2022thesis}.

However, while for $^{76}$Ge and $^{82}$Se a good agreement between theoretical and experimental data was obtained using the effective operator \cite{coraggio2019renormalization}, in the present case, even in the region 0$-$6.6 MeV where agreement is expected, we overestimate the experimental $\Sigma\it{B}$(GT). From experiment, $\Sigma\it{B}$(GT) are 0.085(3) and 0.072(3) for isomeric and ground states, respectively, whereas they are 0.1182 and 0.1466, respectively, from theory. These differences might be contributed to by experimental limits due to some unobserved states. However, given high statistics of $^{80m}$Ga and $^{80g}$Ga, 0.975(25) $\times$10$^8$ and 1.24(3) $\times$10$^8$, respectively, the higher detection efficiency in the low-energy region, and being far away from $\it{S}_n$ (8.08 MeV), i.e., below the continuum region, the $\beta$ intensities of unobserved states must be very small. Therefore, they would be populated via forbidden $\beta$ decay rather than allowed GT transitions. Consequently, we carefully claim that these differences are mainly caused by overestimates of $\it{B}$(GT)'s from theory. This reveals the challenge in precisely reproducing the $\beta$ strength, especially in the low but important energy region with fine structure located within the $\it{Q}_{\beta}$ window. Consequently, further renormalization of the shell-model effective GT operator is needed, probably via adding a two-body term \cite{gysbers2019discrepancy,PhysRevC.109.014301} in \textbf{$\Sigma\sigma{\tau}^-$} operators.

For $\beta$- and $\gamma$-decay pattern analyses, due to lack of structure information of the 4$_1^+$ state and candidate 4$_2^+$ state in $^{80}$Ge, only decay patterns of precursor and $\beta$-populated states to the 2$_1^+$ and 2$_2^+$ states are analyzed. The 2$_1^+$ state has a richer multicorrelation and larger quadrupole deformation than the 2$_2^+$ state, representing higher collective excitation than the 2$_2^+$ state \cite{iwasaki2008persistence,rhodes2022evolution}. This characteristic was confirmed by analysis of the neutron and proton components in reproducing the $\it{B}$($\it{E}$2;0$_1^+$ $\rightarrow$ 2$_1^+$) measurement, $\it{A}_n$ ($\textless$2$_1^+\parallel\it{E2}\parallel$0$_1^+\textgreater_n$) = 13.3 $\it{e}$ fm$^2$ and $\it{A}_p$ ($\textless$2$_1^+\parallel\it{E2}\parallel$0$_1^+\textgreater_p$) = 17.5 $\it{e}$ fm$^2$, respectively, while $\it{A}_n$ = 0 $\it{e}$ fm$^2$ for the 2$_1^+$ state in $^{82}$Ge \cite{padilla2005b}. Consequently, if a precursor in $^{80}$Ga or a high-lying state in $^{80}$Ge has a large collectivity, it will have larger probability of $\beta$ decaying or $\gamma$ deexciting to the 2$_1^+$ state than to the 2$_2^+$ state.

\begin{figure}[!htb]
\centering
\includegraphics[width=1.0\columnwidth]{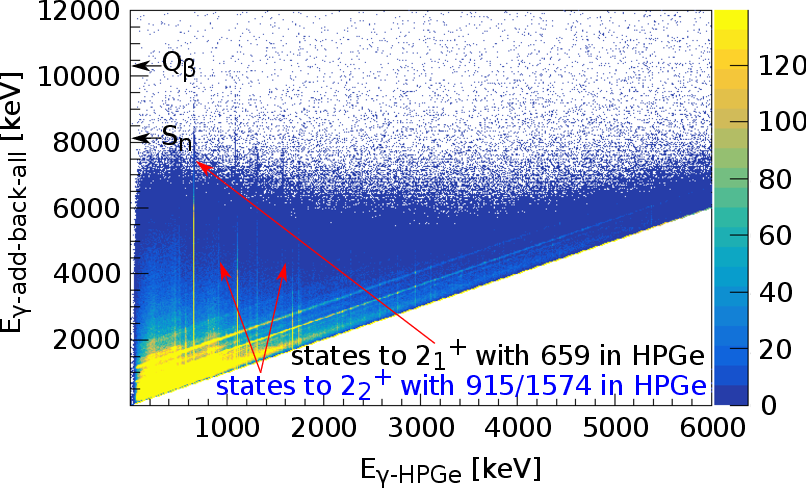}
\caption{$\gamma$-$\gamma$ matrix filled by energies of add-back-all $\gamma$ rays from all detectors (in PARIS clusters, NaI crystal works as veto detector) and single $\gamma$ rays from HPGe detectors. $\it{X}$-axis values of marked lines are 659.2, 915.1, and 1573.6 keV, respectively.}
\label{fig2_1}
\end{figure}

\begin{figure*}[!htb]
\centering
\includegraphics[width=1.0\textwidth]{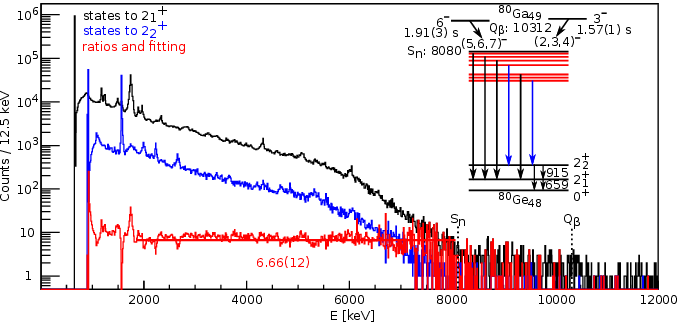}
\caption{$\beta$-gated $\gamma$ spectra measured with add-back-all mode. Black: gate on 659.2 keV $\gamma$ ray; blue: gate on 915.1 or 1573.6 keV $\gamma$ ray; red: ratios of black and blue histograms. Inset: simplified level scheme showing more $\beta$-populated states from $^{80m}$Ga deexciting to the 2$_1^+$ state than to the 2$_2^+$ state in $^{80}$Ge.}
\label{fig2}
\end{figure*}

Figure $\ref{fig2_1}$ shows the $\gamma$-$\gamma$ matrix filled by energies of add-back-all $\gamma$ rays from all detectors and single $\gamma$ rays from HPGe detectors with a coincidence time window of 50 ns. For a better understanding of this, Fig. $\ref{fig2}$ presents the $\beta$-gated $\gamma$ spectra measured with add-back-all mode, as shown in Fig. $\ref{fig2_1}$. In Fig. $\ref{fig2}$ energies were summed up between two HPGe detectors and twenty-seven PARIS phoswiches but on conditions of observation of the 659.2 keV $\gamma$ ray from the 2$_1^+$ state and of the 915.1 keV or 1573.6 keV $\gamma$ ray from the 2$_2^+$ state in a HPGe detector, respectively. Therefore, these spectra present the $\beta$-populated excitations in $^{80}$Ge which deexcite to the above-mentioned two low-lying states, as illustrated by the simplified scheme in the inset of Fig. $\ref{fig2}$. Note that above 5.6 MeV only $\beta$-delayed $\gamma$ rays from $^{80}$Ge exist in the spectra since the $\it{Q}_{\beta}$ values of the daughter nuclei of $^{80}$Ge and $^{80}$As are 2679(4) and 5545(4) keV, respectively \cite{wang2017ame2016}. Furthermore, since the beam was pure and anti-coincidence technique was applied via the outer-layer large NaI crystals in PARIS clusters, the spectra are background free in the high-energy part. Random pileup events were also removed thanks to the pulse-shape discrimination function of the PARIS clusters. This is clear from the clean background above the $\it{Q}_{\beta}$ region where the statistics are constant and very low. The red curve in Fig. $\ref{fig2}$ presents the statistical ratios of these two histograms, where the solid line is a constant fitting between 1.8 MeV and $\it{S}_n$. One can observe that the ratios of these two histograms are approximately constant in a wide energy range. The value is 6.66(12). This is consistent with the parallel profiles of black and blue spectra in a global view except for some fluctuations like the peak at 4324.2 keV.

\begin{figure}[!htb]
\centering
\includegraphics[width=0.9\columnwidth]{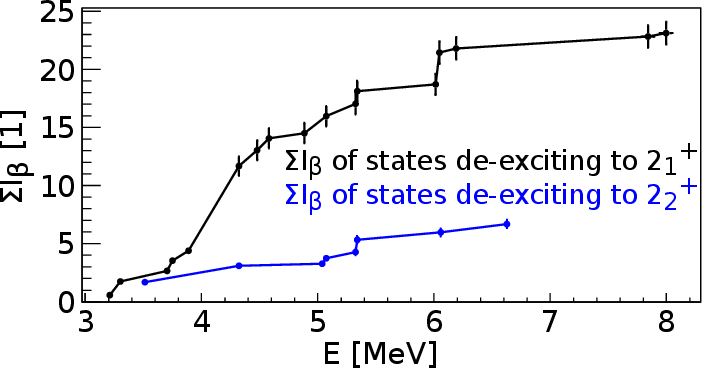}
\caption{Black: cumulative $\it{I}_{\beta}$ of states above 2 MeV $\beta$ populated by $^{80m}$Ga which deexcite to the 2$_1^+$ state. Blue: cumulative $\it{I}_{\beta}$ of states, $\beta$-populated by $^{80m}$Ga, deexciting to the 2$_2^+$ state.}
\label{fig4}
\end{figure}

For the spectrum ratio, since the two spectra are specifically $\gamma$-line gated, the ratios between them include two contributions: (1) detection efficiencies of $\gamma$ lines on which we put gates. [efficiencies of HPGe detectors for 659.2, 915.1 and 1573.6 keV $\gamma$ lines are 0.0395(4), 0.030(3) and 0.018(1), respectively]; (2) the 2$_2^+$ state deexciting to 2$_1^+$. The $\it{I}_{\gamma}$ ratio between 915.1 and 1573.6 keV is 1.238(48). In order to not to distort the spectrum ratio, we do not manually subtract the 915-keV gated spectrum from the black curve as the detection efficiencies of the 659 and 915 keV $\gamma$ lines are different.

In order to obtain an accurate ratio of states deexciting to 2$_1^+$ and 2$_2^+$ states, populated by $^{80m}$Ga, cumulative $\it{I}_{\beta}$ are analyzed, as shown in Fig. $\ref{fig4}$. $\Sigma\it{I}_{\beta}$ of states above 2 MeV which deexcite to the 2$_1^+$ state is 23.1(10)$\%$ whereas it is 6.7(3)$\%$ for states deexciting to the 2$_2^+$ state. Regarding common states, as listed in Table \ref{tab2}, $\it{I}_{\beta}$'s were separated according to $\gamma$ branching ratios, R$_{{\gamma}{-}(2_1^+/2_2^+)}$. Notably, for the levels which have only two depopulating paths the uncertainties in the two branches are not identical, as they are calculated independently using individual deexciting $\gamma$ information. The ratio of $\Sigma\it{I}_{\beta}$ is 3.45(23). Therefore, the spectrum ratio and the $\Sigma\it{I}_{\beta}$ ratio both evidence that more states with $\gamma$ connecting to the 2$_1^+$ state are populated in $\beta$ decay of $^{80m}$Ga. Simultaneously, $\gamma$ decay to 2$_2^+$ state results in a significant percentage/competition as well. Note that, besides influence of structure, 3.83(2) includes contribution of the factor due to the energy-gap difference between high-lying states to low-lying 2$_1^+$ and 2$_2^+$ states \cite{hamilton1975electromagnetic}. For a 7 MeV state, the factor is 1.5959(4) and 2.179(1) for $\it{E}$1/$\it{M}$1 and $\it{E}$2/$\it{M}$2 transitions, respectively. The realistic shell-model calculation supports this conclusion, in which the quadrupole moments ($\it{Q}_2^0$) of each state in Figs. $\ref{fig1}$(a) and $\ref{fig1}$(b) obtained with the effective operators were calculated. 52.1$\%$ of 3$^-$ $\beta$-populated and 83.6$\%$ of 6$^-$ $\beta$-populated states have $\it{Q}_2^0$ values which are larger than 10 $\it{e}$ fm$^2$.

\begin{table}[htbp]
\centering
\caption{Competitions between deexcitation to 2$_1^+$ and to 2$_2^+$ states for $^{80m}$Ga 3$^-$ $\beta$-populated states.}  \vspace{0.5mm}
\label{tab2}
\renewcommand\arraystretch{0.7}
\setlength{\tabcolsep}{7pt}
\begin{tabular*}{0.9\linewidth}{lllll}
E$_\textnormal{levels}$ (keV) &4324.2 &5072.1 &5324.5 &5338.1	\\
\hline
\hline
Br to 2$_1^+$ &50(1)$\%$ &76(4)$\%$ &66(7)$\%$ &51(4)$\%$ 	\\
Br to 2$_2^+$ &9.6(5)$\%$ &24(3)$\%$ &34(4)$\%$ &49(5)$\%$ 		\\
R$_{\gamma}$-$(2_1^+/2_2^+)$ & 5.2(3)	& 3.1(4)  &1.9(3) & 1.1(1)	\\

\end{tabular*}
\end{table}

\begin{figure}[!htb]
\centering
\includegraphics[width=1.0\columnwidth]{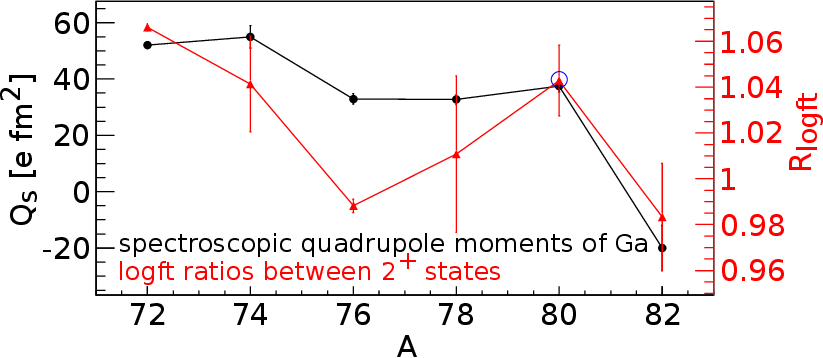}
\caption{Black: spectroscopic quadrupole moments of $^{72g,74g,76g,78g,80m,82g}$Ga \cite{PhysRevA.6.1702,STONE200575,PhysRevC.84.024303,STONE20161,STONE2021,PhysRevC.82.051302,PhysRevC.96.044324}. Red: log $\it{ft}$ ratios of 2$_2^+$ and 2$_1^+$ states in $^{72,74,76,78,80,82}$Ge \cite{KRANE20121649,CAMP1968561,RESTER1971461,Taylor1975,CAMP1971145,silwal2018detailed,PhysRevC.22.2178,PhysRevC.93.044325}.}
\label{fig3}
\end{figure}

For further investigation of the effect of nuclear collectivity on the $\beta$-decay property, we analyze the correlation between quadrupole deformation of precursors and relative $\beta$-transition strength to highly quadrupole deformed 2$_1^+$ (2$_2^+$ for $^{72,82}$Ge) states in germanium isotopes. Less deformed 2$_2^+$ (2$_1^+$ for $^{72,82}$Ge) states were selected as references. It is worth noting that 2$_2^+$ states, as a band head of the quasi-$\gamma$ band in $^{72}$Ge \cite{AYANGEAKAA2016254} and as a rotational state in $^{82}$Ge \cite{PhysRevC.84.024305}, have larger quadrupole deformations than the 2$_1^+$ states of the ground-state bands which are usually thought to have spherical shapes for nuclei with magic numbers of neutrons and/or protons, e.g., $\it{N}$ = 40 and $\it{N}$ = 50 in $^{72}$Ge and $^{82}$Ge, respectively. In Fig. $\ref{fig3}$, the black curve presents spectroscopic quadrupole moments ($\it{Q}_s$) of $^{72g,74g,76g,78g,80m,82g}$Ga \cite{PhysRevA.6.1702,STONE200575,PhysRevC.84.024303,STONE20161,STONE2021,PhysRevC.82.051302,PhysRevC.96.044324} while the red one shows the ratios of log $\it{ft}$ values between 2$_2^+$ and 2$_1^+$ in $^{74,76,78,80}$Ge and between 2$_1^+$ and 2$_2^+$ in $^{72,82}$Ge; R$_{\textnormal{log}{\hspace{0.035in}}\it{ft}}$ values represent log $\it{ft}$ ratios between small and large deformed 2$^+$ states. The data point surrounded with a blue circle is from this work while others are from Refs. \cite{KRANE20121649,CAMP1968561,RESTER1971461,Taylor1975,CAMP1971145,silwal2018detailed,PhysRevC.22.2178,PhysRevC.93.044325}. The increase of R$_{\textnormal{log}{\hspace{0.035in}}\it{ft}}$ indicates that the highly deformed 2$^+$ state becomes more favored than the less deformed 2$^+$ state, and vice versa. Therefore, if nuclear collectivity has significant impact on the $\beta$-decay property, the precursor with larger $\it{Q}_s$ should have a larger R$_{\textnormal{log}{\hspace{0.035in}}\it{ft}}$. Clearly, this positive correlation has been built and is observable, as shown in Fig. $\ref{fig3}$, from $\it{A}$ = 72 to $\it{A}$ = 82. Given nonequal absolute collectivities of 2$^+$ states in the presented Ge isotopes, one should pay attention to the tendencies of the two curves rather than the comparison of the absolute values between isotopes. Regarding the influences of precursors' spin-parity on this correlation, we note that since all involved states of Ga isotopes have (2,3)$^-$ assignments except $^{78g}$Ga that has (3$^+$) assignment, they all populate 2$_{1,2}^+$ via first-forbidden $\beta$ transitions. Therefore, we did not find significant influences of precursors' spin in our study including $^{78g}$Ga. However, in the case of other isotopes, if the spin gaps between precursors and analyzed 2$_{1,2}^+$ states in daughter nuclei are 2 or larger and their parity is different, e.g., first-forbidden unique or second-forbidden unique transitions, the influences of spin and parity on this correlation might appear.

\section{IV. Summary and conclusions}

We have presented here experimental evidence in $^{80}$Ga for simultaneous impacts of nuclear shell structure that lead to jump structure in $\Sigma\it{B}$(GT), and collectivity which results in positive correlation between deformation of precursor and selectivity to highly deformed 2$_1^+$ state, on the $\beta$-decay property. To further justify this conclusion, we have performed a realistic shell-model calculation with the effective Hamiltonian and GT operator derived consistently within the framework of the many-body perturbation theory. Though remarkable, the renormalization is insufficient to reproduce the experimental strength, probably evidencing the need of a two-body term for the GT operator.

\section{Acknowledgment}

The authors thankfully acknowledge the work of the ALTO technical staff for the excellent operation of the ISOL source. R.L. acknowledges support by China Scholarship Council under Grant No.201804910509 and a KU Leuven postdoctoral fellow scholarship. C.D., A.K., and L.A.A. have received funding from European Union's Horizon 2020 research and innovation program under Grant Agreement No. 771036 (ERC CoG MAIDEN). Use of the PARIS modular array from the PARIS Collaboration and Ge detectors from the French-UK IN2P3-STFC Gamma Loan Pool are acknowledged.

\bibliography{References}
\onecolumngrid
\end{document}